\documentclass[11pt,a4paper]{article}

\usepackage[]{geometry}  
\usepackage[T1]{fontenc}  
\usepackage[utf8x]{inputenc}  
\usepackage{lineno}  
\usepackage{fancyhdr}  
\usepackage{enumitem}  
\usepackage{hyperref}  
\usepackage{titling}  
\usepackage{natbib}  
\usepackage{mathtools}  
\usepackage{titlesec}  
\usepackage{lastpage}  

\usepackage[english]{babel}  
\usepackage{amsmath}  
\usepackage{amsfonts}  
\usepackage{amssymb}  
\usepackage{wasysym}  
\usepackage{bbm}  
\usepackage{array}  
\usepackage{xr}  
\usepackage{verbatim} 
\usepackage{float} 

\usepackage{svg}
\usepackage{multicol}
\usepackage{caption}

\hypersetup{%
  pdfborderstyle={/S/U/W 1}
}

\setcitestyle{round,numbers}
\titleformat{name=\section}{\normalfont\Large\bfseries}{\thesection}{1em}{}
\titleformat{name=\subsection}{\normalfont\large\bfseries}{\thesubsection}{1em}{}
\titleformat{name=\subsubsection}{\normalfont\normalsize\bfseries}{\thesubsubsection}{1em}{}

\newcommand{\email}[1]{\href{mailto:{#1}}{{#1}}}

\newcommand{\keywords}[1]{\textbf{Keywords}: {#1}}

\newcommand{\optincludegraphics}[2][]{}

\newcommand{\optinput}[1]{}

\newtagform{brackets}{[}{]}
\usetagform{brackets}

\newcommand{\thejournal}[1]{Magnetic Resonance in Medicine}

\title{MR Simulation with Phase Distribution Graphs: Off-Resonant Pulse Response and Slice-Selection}

\begin{document}

{\noindent\LARGE\bf \thetitle}

\bigskip

\begin{flushleft}\large
	Felix Dietz\textsuperscript{1},
	Jonathan Endres\textsuperscript{1},
	Simon Weinmüller\textsuperscript{1},
    Johanna Link\textsuperscript{1}, 
	Moritz Zaiss\textsuperscript{1,2}
\end{flushleft}

\noindent
\begin{enumerate}[label=\arabic*]
\item Department of Neuroradiology, University Hospital Erlangen, Erlangen, Germany
\item Department of Artificial Intelligence in Biomedical Imaging (AIBE), \\Friedrich-Alexander University Erlangen-Nürnberg, Erlangen, Germany
\end{enumerate}

\begin{center}
\email{felix.dietz@fau.de}
\end{center}

\bigskip


\vfill






\begin{abstract}
\textbf{Purpose}: Phase Distribution Graphs were introduced as a powerful tool for on-resonant MRI simulations. Herein, we extend the concept beyond the assumptions of hard and on-resonant RF pulses in Phase Distribution Graph simulations, while retaining differentiability across all parameters. The extension into the frequency domain further generalizes the Phase Distribution Graph simulation framework for sequence simulation and optimization.

\textbf{Theory and Methods}:  Using an effective axis rotation model, the RF operator is generalized to off-resonant rotations, originating from field inhomogeneities, chemical shift, and field gradients. Discretization of the RF pulse shape enables the modeling of the magnetization response to shaped RF pulses and thus slice selection.

\textbf{Results}: Including off-resonant pulse response into a PDG based MRI simulation enables the simulation of key MRI features, such as fat suppression, binomial pulses, or Bloch-Siegert shifts. The discretization of RF shapes enables the simulation of their respective frequency response and thus slice selective excitation, refocusing or saturation pulses, including multi-band excitation. In combination with flow, typical inflow artifacts in slice-selective MRI can be simulated.  

\textbf{Discussion and Conclusion}:
Extension for off-resonance removes a major limitation of phase distribution graphs and makes the simulation framework more realistic.
As the method is not restricted to specific RF shapes, optimization of RF pulse shapes in amplitude, phase and frequency is now possible in the context of or jointly with specific MR sequences.

\end{abstract}

\bigskip
\keywords{MRI simulation, Bloch, extended phase graph, slice-selective RF pulses, off-resonant pulses}


\begin{multicols}{2}
\section{Introduction}

Magnetic resonance imaging (MRI) is governed by the interaction of radiofrequency (RF) pulses, gradient fields and the resulting spin dynamics. Accurate simulation of these processes is essential for sequence design and optimization. Some key MRI features, including chemical-shift-based fat–water separation, off-resonance effects caused by field inhomogeneities and slice-selective excitation, require the modeling of the magnetization response to off-resonant pulses.\\

\noindent
A conventional approach to MRI simulation is the execution of the Bloch equations for a large number of spins, referred to as isochromats. Although isochromat simulations are widely applicable, these approaches require dense spin sampling, resulting in substantial computational cost. In contrast, the Extended Phase Graph (EPG) formalism \cite{hennig_multiecho_2003,weigel_extended_2015} provides a compact representation of magnetization by tracking discrete coherence states. Extensions such as slice-selective EPG \cite{Ostenson2020} and a spinor-based formulation of EPG \cite{Li2024SpinorEPG} have expanded its scope, but rely on assumptions regarding sequence structure or periodicity, hindering the simulation of general sequence types.

\noindent
Phase Distribution Graph (PDG) simulations \cite{Endres2024}, introduced within the MR-zero framework \cite{Loktyushin2021,mrzero_framework}, provide a computationally efficient and end-to-end differentiable alternative without imposing such restrictions. However, the current PDG formulation is limited to on-resonant and instantaneous RF pulses.\\ 

\noindent
We extend the PDG framework with a unified operator-based treatment of RF–magnetization interactions that incorporates frequency-selective excitation and allows for the simulation of soft RF pulses. Initial limitations of the PDG framework are mitigated, providing a generalized phase graph-based MRI simulation framework. The resulting framework retains differentiability in all parameters, enabling efficient simulation and optimization of MRI sequences with respect to frequency and slice-profile dependent effects.

\section{Theory}

For the Phase Distribution Graph formulation of the MRI process magnetization is separated into configuration states using a complex coordinate system $M_{+-z}$, defined via the unitary transformation $U$:
\begin{equation}
    M_{+-z} = U \cdot M_{xyz} 
\end{equation}
\begin{equation}
\label{eq:UnitaryTransfromM}
    U =  \begin{pmatrix}
       1 & +i & 0 \\
       1 & -i & 0 \\
       0 & 0 & \sqrt{2} 
    \end{pmatrix}
\end{equation}
where $M_{xyz}$ denotes the magnetization in the rotating reference frame (RRF) fixed to the water resonance. 

\subsection*{Radiofrequency pulses}
Radiofrequency (RF) pulses are described using the hard pulse approximation, as an instantaneous rotation around an axis $(\cos \phi, \sin \phi, 0)$ where $\phi$ is the RF phase. The operator is constructed from elementary rotations around the axis of the RRF. In order to include the off-resonant pulse response, the RF operator needs to be derived, including the tilt angle $\Delta = \arctan(\Delta\omega / \omega_1)$ of the RF rotation axis in the RRF. Essentially, an off-resonant RF pulse rotates magnetization around an axis $(\cos \phi \cos \Delta, \sin \phi \cos \Delta, \sin \Delta)$ by an effective angle $  \alpha_\text{eff} = \alpha \cdot \sqrt{1+ \left( \Delta \omega/\omega_1 \right)^2} $. The full operator including off-resonances is constructed as
\begin{equation}
\label{eq:OffResOperatorConstruction}
\begin{aligned}
     RF(\alpha_\text{eff}, \phi, \Delta) = U R_z(\phi)&R_y(\Delta)R_x(\alpha)  \\
     \cdot &R_y(-\Delta)R_z(-\phi) U^{-1}
\end{aligned}
\end{equation}
where $R_{\hat{e}}$ with $\hat{e} \in \{x,y,z\}$ are the three dimensional rotation matrices around axis $\hat{e}_i$ and U is the unitary transformation matrix from \autoref{eq:UnitaryTransfromM}.

\noindent
This derivation is performed in the frame rotating at the RF carrier frequency. However, PDG states are represented in a frame rotating at the water resonance frequency. Consequently, an additional phase accumulation, dependent on the RF duration $\tau_p$, occurs  $\phi_\text{off} = \Delta\omega \cdot \tau_p$. This phase evolution is incorporated through an additional rotation about the z-axis.

\subsection{Spectral and spatially selective excitation}
The field inhomogeneities, the chemical shift and slice-selection gradients modify the local Larmor frequency and can therefore be modeled as spatially varying frequency offsets. The resulting off-resonance frequency is given by 
\begin{equation}
\Delta\omega(\mathbf r)
= 2\pi \left[\Delta f_{\mathrm{RF}}-\Delta f_0(\vec{r}) -\delta(\mathbf r)-B_g(\mathbf r)\right]
\label{eq:LarmorLocal}
\end{equation}
where $\Delta f_{\mathrm{RF}}$ denotes the RF pulse frequency offset, $\Delta f_0(\vec{r})$ inhomogeneities of the $B_0$ field, $\delta(\mathbf r)$ the chemical-shift-induced frequency shift, and $B_g(\mathbf r)$ the frequency offset generated by the slice-selection gradient. 

\noindent
In phase-graph-based simulations, RF pulses and gradients are treated as separate events. The effect of the slice-selection gradient is therefore incorporated through the spatially dependent frequency offset in \autoref{eq:LarmorLocal}, which directly determines the off-resonance tilt angle $\Delta$ used in the pulse operator.

\subsection*{Soft RF pulses}
To model arbitrary RF waveforms within the PDG framework, a shaped RF pulse is discretized into $N$ equal-duration sub-pulses. This approximation captures both the RF pulse shape and magnetization evolution during excitation. The flip angle of the $n$-th sub-pulse is given by

\begin{equation}
\alpha_n = \int_{t_n}^{t_{n+1}} \omega_1(t)\,\mathrm{d}t .
\end{equation}

\subsection*{Magnetization dynamics during RF irradiation}

The state dynamics simultaneously including RF irradiation, gradient field $G$, and $T_1$ and $T_2$ relaxation are given by the Bloch equations in the Fourier domain.
Exploiting the state partitioning of the PDG framework, for sufficiently short discretization $\Delta t$, the dynamics of existing states 
can be separated from the recovery of $z$-magnetization, which creates new and undephased $z$-magnetization states. 
The dynamics of the existing states can then be decomposed into
\begin{equation}
\frac{\partial \vec{F}(\vec{k},t)}{\partial t}
=
(T+RF)\vec{F}(\vec{k},t),
\end{equation}
where 
\begin{equation}
    T = \mathrm{diag} \left(\frac{\gamma}{2\pi}G\frac{\partial}{\partial k_z}-\frac{1}{T_2}, -\frac{\gamma}{2\pi}G\frac{\partial}{\partial k_z}-\frac{1}{T_2}, -\frac{1}{T_1}  \right)
\end{equation} 
contains the gradient and relaxation terms and 
\begin{equation}
    RF = \begin{pmatrix}
        0 & 0 & \frac{i}{\sqrt{2}}\omega_1 e^{i\phi} \\
        0 & 0 & -\frac{i}{\sqrt{2}}\omega_1 e^{-i\phi} \\
        \frac{i}{\sqrt{2}} \omega_1 e^{-i\phi} & -\frac{i}{\sqrt{2}} \omega_1 e^{i\phi} & 0
    \end{pmatrix}  
\end{equation}
describes RF-induced state transitions. Since $T$ and $RF$ do not commute, time evolution over a step $\Delta t$ is approximated using second-order Strang splitting:
\begin{equation}
\vec{F}(\vec{k}, t+\Delta t)
\approx
e^{T\Delta t/2}
e^{RF\Delta t}
e^{T\Delta t/2}
\vec{F}(\vec{k}, t).
\end{equation}
The matrix exponential $e^{T\Delta t/2}$ contains the relaxation and gradient shift operators for half the step duration $\Delta t/2$. $e^{RF\Delta t}$ results in the RF description given by the operator form \autoref{eq:OffResOperatorConstruction} for the whole simulation time step $\Delta t$.

\section{Methods}

\subsection*{Simulation setup}

The extended PDG simulation is built upon the original implementation in Python/PyTorch, which provides CUDA acceleration. The prepass uses a Rust backend. Simulations were run on the GPU. 

\noindent
Virtual phantoms were built from the data provided by the BrainWeb Data base \cite{collins1998brainweb}, filled with realistic values for the tissue parameters at 3T. Fat and water-like fractions were simulated separately when present at the same location, and the respective signals were summed during signal calculation.

\noindent
For simulation, sequences were defined in the Pulseq standard \cite{layton_pulseq_2017}, ensuring direct scanner executability of the sequences used for simulation. The presented experiments focus on validating the developed extension to the PDG framework and are described in the following. Unless otherwise specified, the base sequence for all imaging experiments is a FLASH sequence (FA: $10^\circ$, TR: $4 \, \mathrm{ms}$, TE: $2 \, \mathrm{ms}$, in-plane resolution: $64\times64$, FOV: $200 \, \mathrm{mm}$), which is adapted based on the application.  

\subsection*{Spectral response}
\subsubsection*{Fat suppression}
The spectral behavior of the RF operator was evaluated using a CHESS pulse fat saturation module \cite{Haase_CHESS} and water-selective excitation realized with a second-order binomial pulse train. Simulation results were compared with a conventional FLASH acquisition. Frequency response spectra were sampled over a dynamic range of $\pm 8 \, \mathrm{ppm}$ to validate the spectral response across a broad range of frequencies.

\subsubsection*{B0 and B1 mapping using off-resonance effect} 
WASABI \cite{WASABI} and Bloch-Siegert \cite{BlochSiegert} field-mapping experiments were simulated to assess longitudinal and transverse magnetization responses of the proposed RF formulation. Both experiments used a FLASH readout, with the WASABI preparation and the binomial pulse excitation added respectively. Reconstructed $\Delta B_0$ and relative $B_1$ maps were compared voxel-wise with the ground truth. Difference maps were computed for quantitative assessment. 

\subsection*{Slice selection}

\subsubsection*{Slice profile simulation}
Slice profiles were simulated using a 1D phantom and acquiring a single line in $k$-space of which a Fast Fourier Transform was computed. The resulting slice profiles are benchmarked with results obtained by numerically solving the Bloch equations corresponding to a single isochromat simulation. As a quantitative comparison metric, relative deviations are computed. 

\subsubsection*{Slice selective excitation}
For slice selective simulations a 3D phantom is used. The magnetization profiles after excitation are evaluated in axial and sagittal views. 

\subsection*{Optimization setup}


\section{Results}
 
\subsection*{Spectral response: fat suppression}
\begin{figure}[h]
    \centering
    \includegraphics[width=0.9\linewidth]{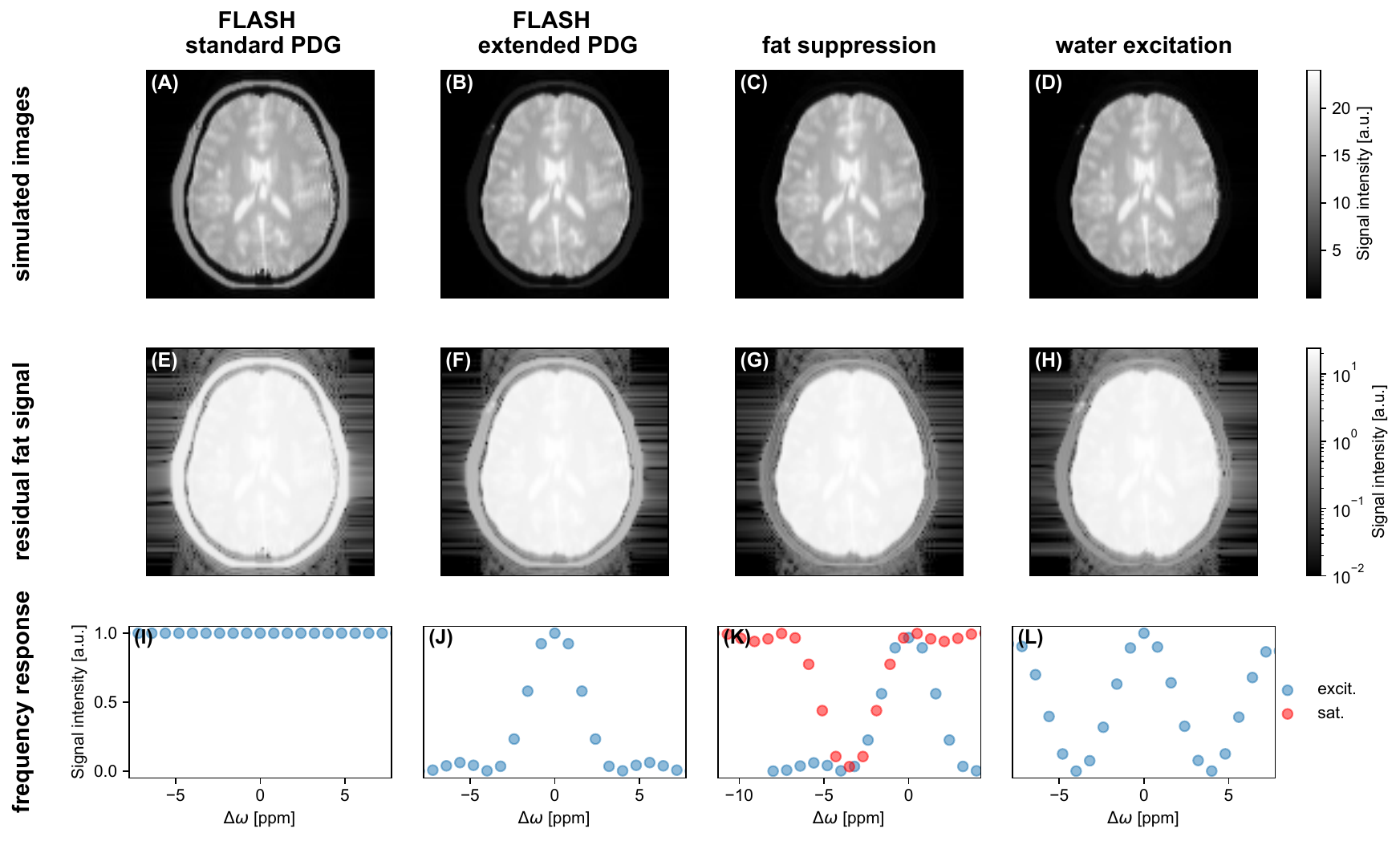}
    \caption{(A) A FLASH readout was used to simulate a reference image, showing the brain surrounded by subcutaneous fat. (B) Simulated fat-suppressed image obtained using the CHESS module. (C) Simulated water-excitation image obtained using a binomial pulse train. (I-L) Frequency response spectra of the methods, centered around their respective center frequency.}
    \label{fig:FatWaterSignalSeparation}
\end{figure}

The spectral selectivity was evaluated on fat–water signal separation, realized by using a CHESS fat-saturation module and a binomial water-excitation pulse train. \autoref{fig:FatWaterSignalSeparation} compares the resulting images with a conventional FLASH acquisition. Compared to the standard PDG in \autoref{fig:FatWaterSignalSeparation} (A,E), the standard rectangular excitation pulse simulated with the off-resonant PDG (B,F) already shows significantly lower fat signal, explained by the frequency response of the excitation pulse (J), which is now correctly modeled. Both fat suppression methods further suppress the fat signal (\autoref{fig:FatWaterSignalSeparation} C,D,G,H). The corresponding frequency response spectra reflected the expected spectral selectivity: fat saturation produced a narrow suppression band centered at the fat resonance, whereas water excitation was centered at the water resonance and exhibited periodic sidebands with spacing $\Delta \omega = 2\pi \Delta f_\text{fat-water}$ due to the temporal spacing of the binomial sub-pulses. These results demonstrate that the proposed off-resonant RF operator accurately reproduces the frequency-selective behavior required for chemical-shift-based excitation and saturation. The spectral response is validated across a broad range of frequencies via the frequency response spectra.

\subsection*{Spectral response: field mapping using off-resonant effects}
\begin{figure}[h]
    \centering
    \includegraphics[width=0.8\linewidth]{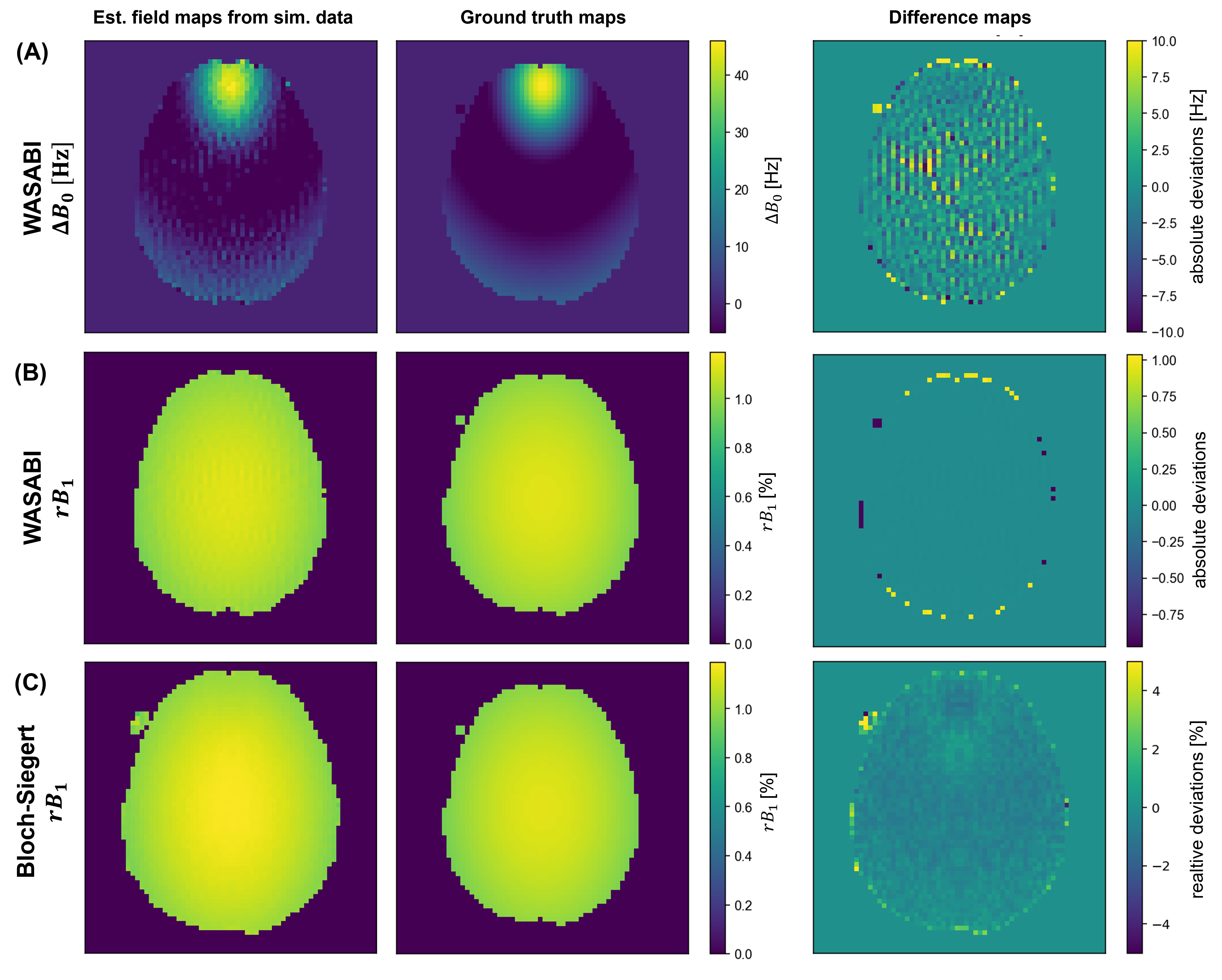}
    \caption{Estimated $\Delta B_0$ (A) and $rB_1$ (B) field maps, obtained by fitting simulated WASABI sequence data, compared with the ground truth maps of the simulation phantom. Difference maps of the $\Delta B_0$ and $rB_1$ field mapping show good agreement between the simulated and the ground truth values, with deviations below $\pm 10 \, \mathrm{Hz}$ for the $\Delta B_0$-map and deviations of the $rB_1$-map appearing solely on the phantom boundaries. Estimated $\Delta B_1$ (C) field maps, obtained by evaluation of simulated Bloch-Siegert sequence data compared with the ground truth map from the simulation phantom. The relative deviation of the obtained Bloch-Siegert map from the ground truth is within a $5\%$ error margin, due to residual $B_0$ contamination.}
    \label{fig:FieldMapping}
\end{figure}

Simulated WASABI and Bloch-Siegert experiments were used to evaluate the response of longitudinal and transverse magnetization to off-resonant RF irradiation. As shown in \autoref{fig:FieldMapping}(A,B), the simulated WASABI-derived $\Delta B_0$ and relative $rB_1$ maps closely matched the ground-truth phantom maps. Deviations in the $\Delta B_0$ estimates remained within approximately $\pm10  \,\mathrm{Hz}$. Deviations in the $rB_1$ maps were observed exclusively at phantom boundaries. Both these effects are likely due to a mismatch in phantom and sequence resolution causing interpolation artifacts. Similarly, the Bloch-Siegert experiment, as shown in \autoref{fig:FieldMapping}(C) accurately reproduced the ground-truth $rB_1$ distribution. The corresponding difference map showed relative errors below $\pm5\%$.

\noindent
Fat-suppression and WASABI cover the near-off-resonant regime, and Bloch-Siegert validates the correctness of the far-off-resonant regime. 
\noindent
Taken together, these results confirm that the extended PDG framework accurately reproduces both longitudinal and transverse magnetization responses, supporting its applicability to quantitative field-mapping experiments.

\subsection{Slice Selection}

\begin{figure}
    \centering
    \includegraphics[width=0.8\linewidth]{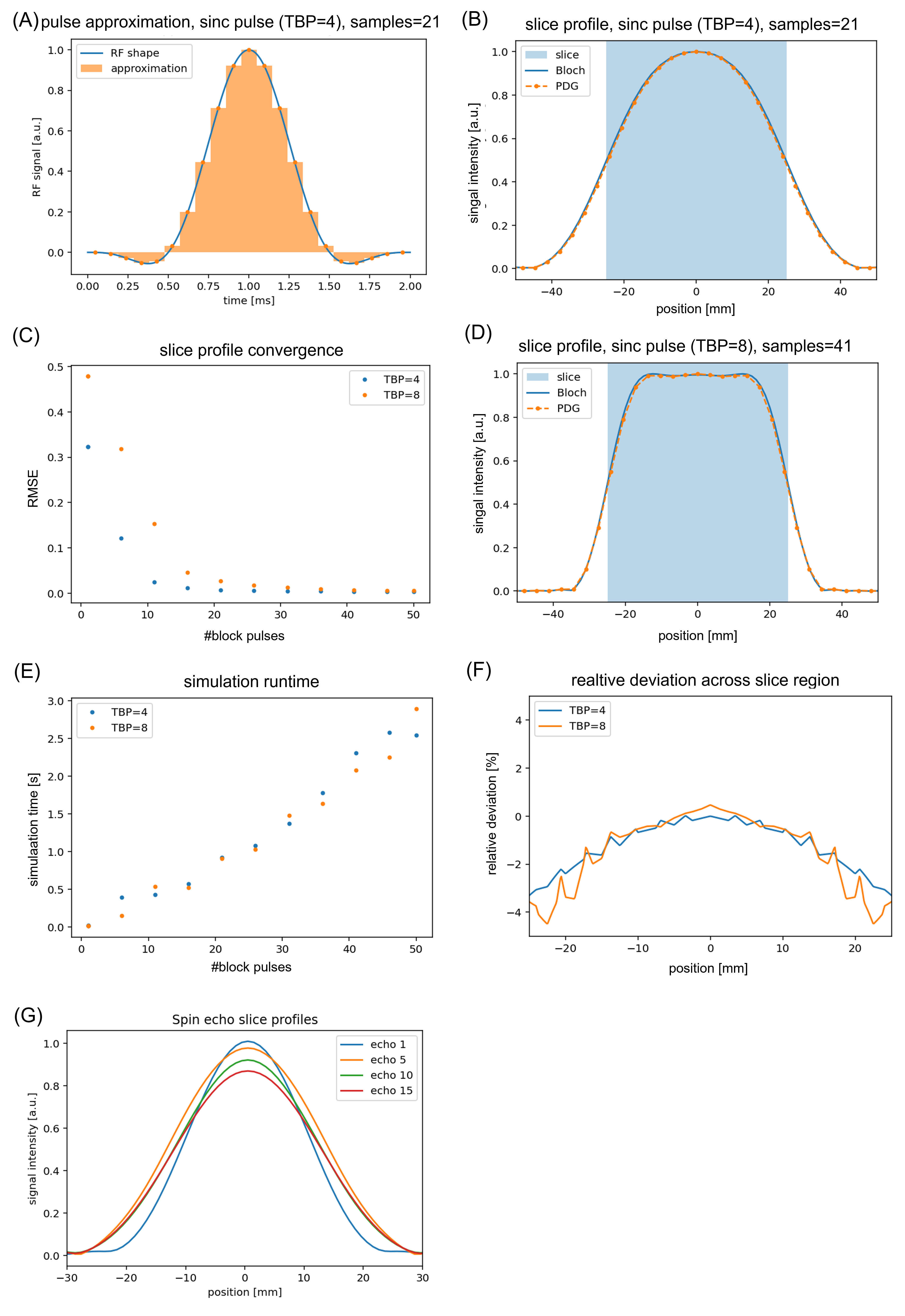}
    \caption{Approximation of a soft RF pulse (sinc-shape, duration 2ms, TBP=4) by a series of 21 block pulses of equal duration (A). Simulated slice profiles for sinc-pulses using 21 (TBP=4) and 41 (TBP=8) block pulses (B,D). Comparison with results from Bloch equation solutions show relative deviations below 5\% (F). PDG simulated slice profiles converge to the expected shape with increasing number of sub-pulses (C). Simulation time increases linearly with the number of samples (E). Slice profile evolution across a spin-echo train (G).}
    \label{fig:ExctProfilesSinc}
\end{figure}

The discretized RF pulse model enables the simulation of slice-selective excitation. \autoref{fig:ExctProfilesSinc} compares PDG-simulated slice profiles for Sinc pulses with time-bandwidth products (TBP) of 4 and 8 against numerical Bloch simulations. Magnitude and phase profiles showed close agreement across the entire slice region. The relative deviations in slice profile magnitude remain below $5\%$ and are minimal at the slice center.  

\noindent
\autoref{fig:ExctProfilesSinc}(C,E) further evaluates convergence and computational cost as a function of pulse discretization. Increasing the number of sub-pulses improved agreement with the Bloch reference. Higher TBP pulses require a finer discretization for accurate representation due to their more complex shape. Simulation time increased linearly with the number of sub-pulses and was independent of pulse shape. 
\noindent
Slice profiles were also simulated for a TSE refocusing train using slice-selective excitation and refocusing pulses (FA: $90^\circ$/$180^\circ$, duration: $1 \, \mathrm{ms}$, Sinc-shape, TBP: $2$, slice thickness: $30 \, \mathrm{mm}$). \autoref{fig:ExctProfilesSinc}(G) shows the simulated slice profiles for several echoes along the echo train. The produced slice profiles widened across the echo train, their $\mathrm{FWHM}$ ranging from $22.0 \, \mathrm{mm}$ to $27.05 \, \mathrm{mm}$. 
\noindent
These results demonstrate that the discretized RF model accurately captures magnetization dynamics during finite-duration RF irradiation.


\subsection*{Viewing slice-selective excitation and simultaneous multi-slice excitation in 3D}
\begin{figure}
    \centering
    \includegraphics[width=0.75\linewidth]{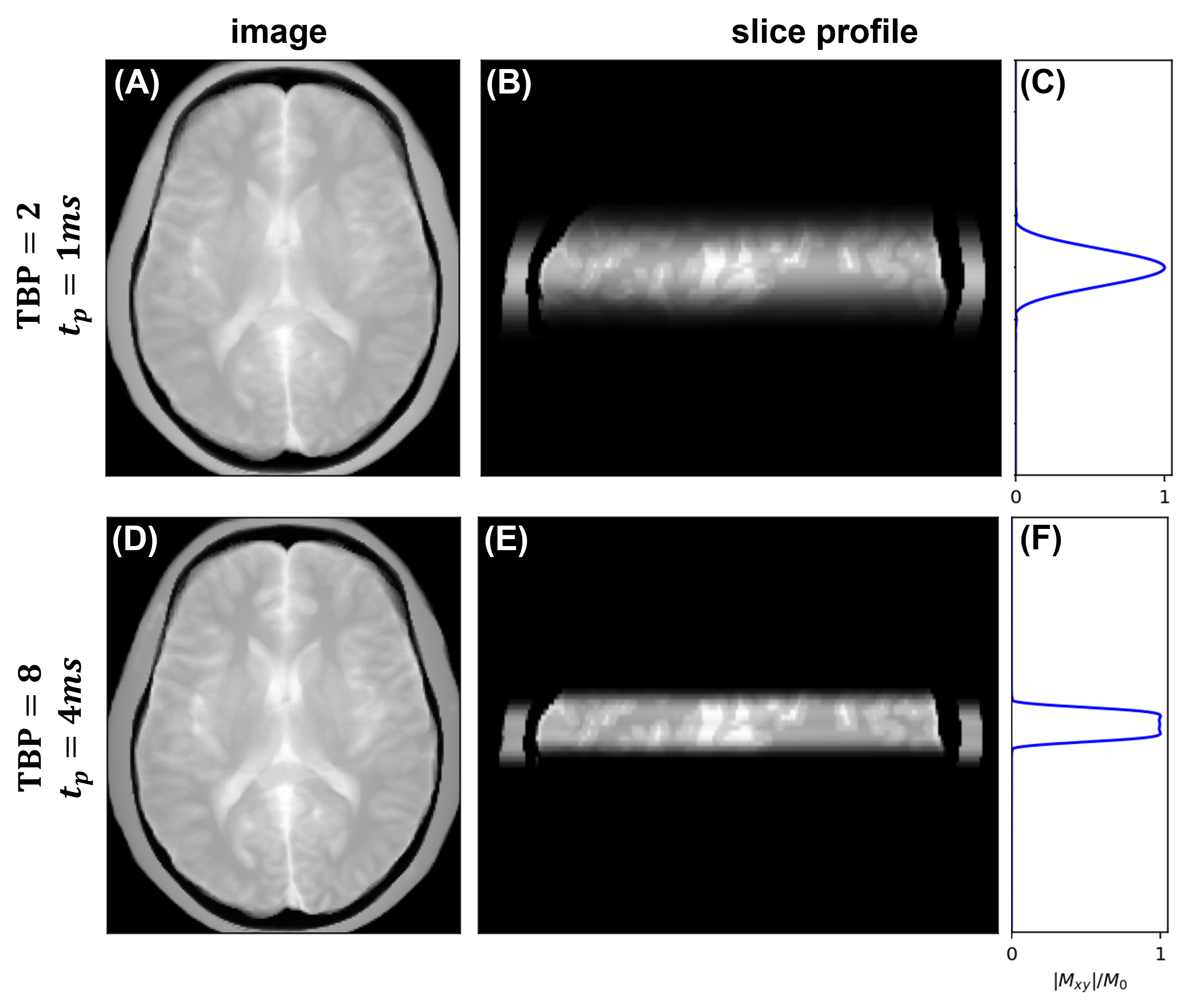}
    \caption{Single slice excitation profiles and reconstructed images.
    Excitation of a $10 \, \mathrm{mm}$ slice is realized using a Sinc pulse with parameters $\mathrm{TBP}=2$, $t_p=1\, \mathrm{ms}$ (A,B) and $\mathrm{TBP}=8$, $t_p=4\, \mathrm{ms}$ (D,E). Substantial blurring is visible in (A) resulting from the broad slice profile. Increased sharpness due to the higher time-bandwidth product (D) requires a longer pulse, increasing the fat shift. Excitation profiles obtained using a Bloch simulator (C,F) are shown for comparison with the PDG simulated slice profiles.}
    \label{fig:sliceSelectiveImaging}
\end{figure}

\begin{figure}
    \centering
    \includegraphics[width=0.75\linewidth]{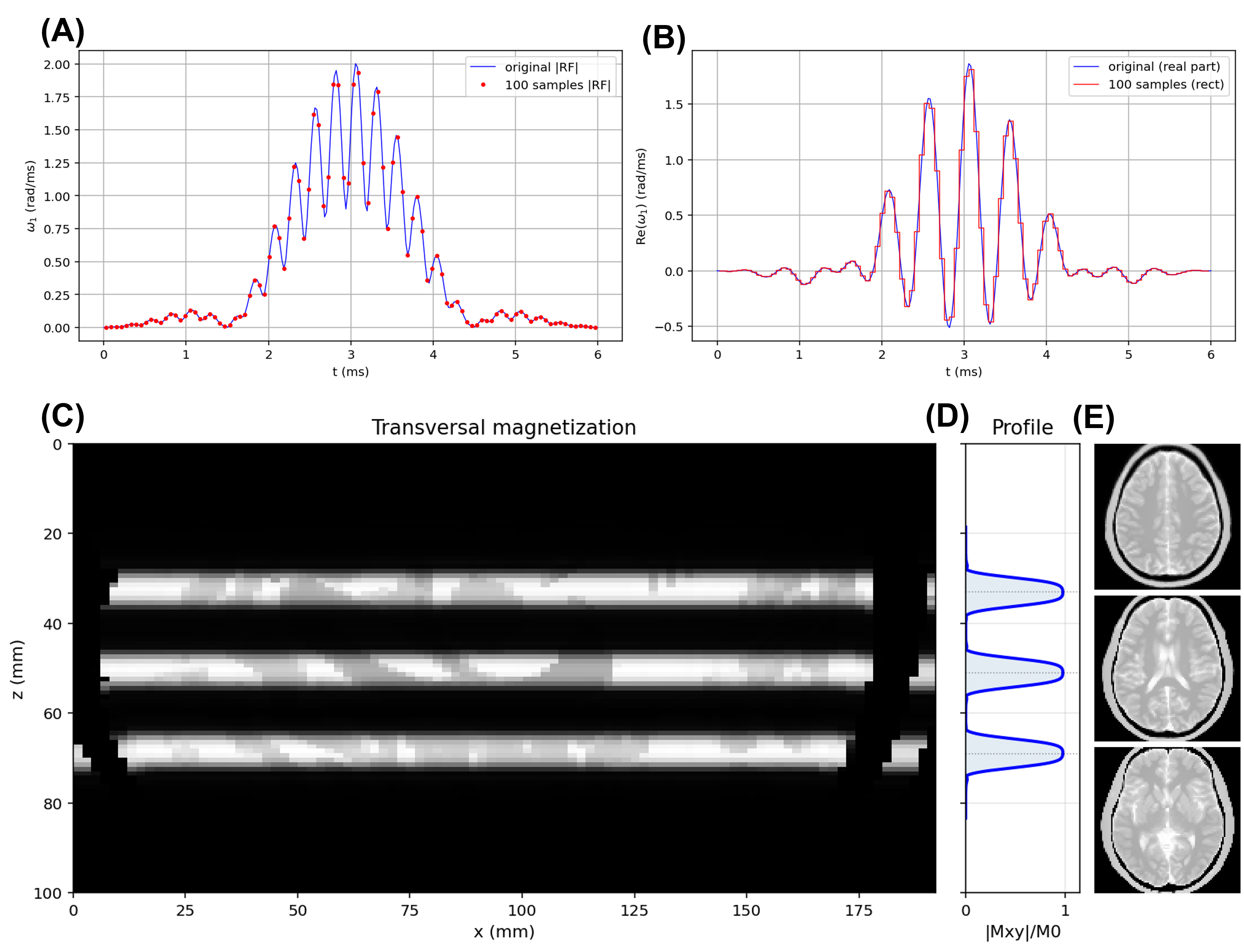}
    \caption{Multi-slice excitation profile and reconstructed images. (A,B) A SMS pulse is split into 100 sub samples. (C) Sagittal view of the transversal magnetization, showing three simultaneously excited slabs matching the (D) slice-selective excitation profile using a Bloch simulator. (E) Summed transversal magnetization of the transversal brain images for each of the three excited slabs.}
    \label{fig:sms}
\end{figure}

Slice selective excitation is simulated on a 3D whole-brain phantom. The resulting magnetization state after the pulse is shown in axial and longitudinal views in \autoref{fig:sliceSelectiveImaging}(A-C). This allows the slice-selection performance of different tissues to be inspected in detail, i.e. the shift in z of the fat excitation is clearly visible in the sagittal views.\\
A 3-band simultaneous multi-slice (SMS) pulse with a slice-select gradient was discretized into 100 sub-pulses (\autoref{fig:sms}(A,B)). The simulated slice profile shows three well-separated excitation bands at the prescribed slice offsets matching the excited transversal magnetization as shown in \autoref{fig:sms}(C,D). 

\subsection*{Flow simulation} 
\begin{figure}
    \centering
    \includegraphics[width=0.66\linewidth]{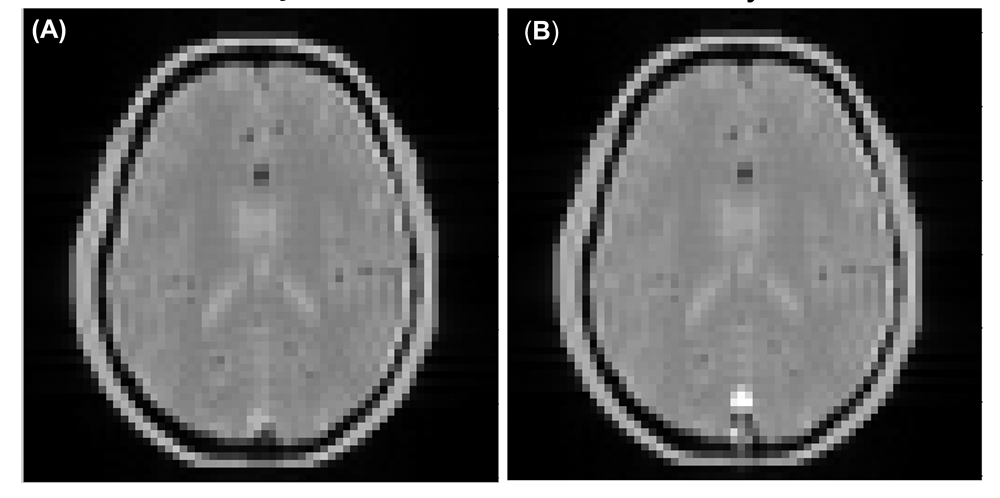}
    \caption{A 64x64 FLASH with 64 dummy pulses using a slice selective excitation pulse is simulated in a 3D brain phantom including a cylindrical flow region in the posterior portion of the sinus sagittalis superior with CSF parameters. Without flow this area is in a low signal steady-state same as the CSF in ventricles (A). With flow fresh magnetization from outside the slice flows into the slice and leads to higher signal intensities in this area (B).}
    \label{fig:flow_in2D_flash}
\end{figure}

\autoref{fig:flow_in2D_flash} demonstrates the simulation of typical in-flow artifacts in slice-selective scans. This capability enables a better understanding of flow-related artifacts and the optimization of mitigation strategies. Furthermore, it establishes a foundation for simulating advanced sequences such as time-of-flight, arterial spin labeling, and perfusion.

\subsection*{Optimization}
\begin{figure}
    \centering
    \includegraphics[width=0.8\linewidth]{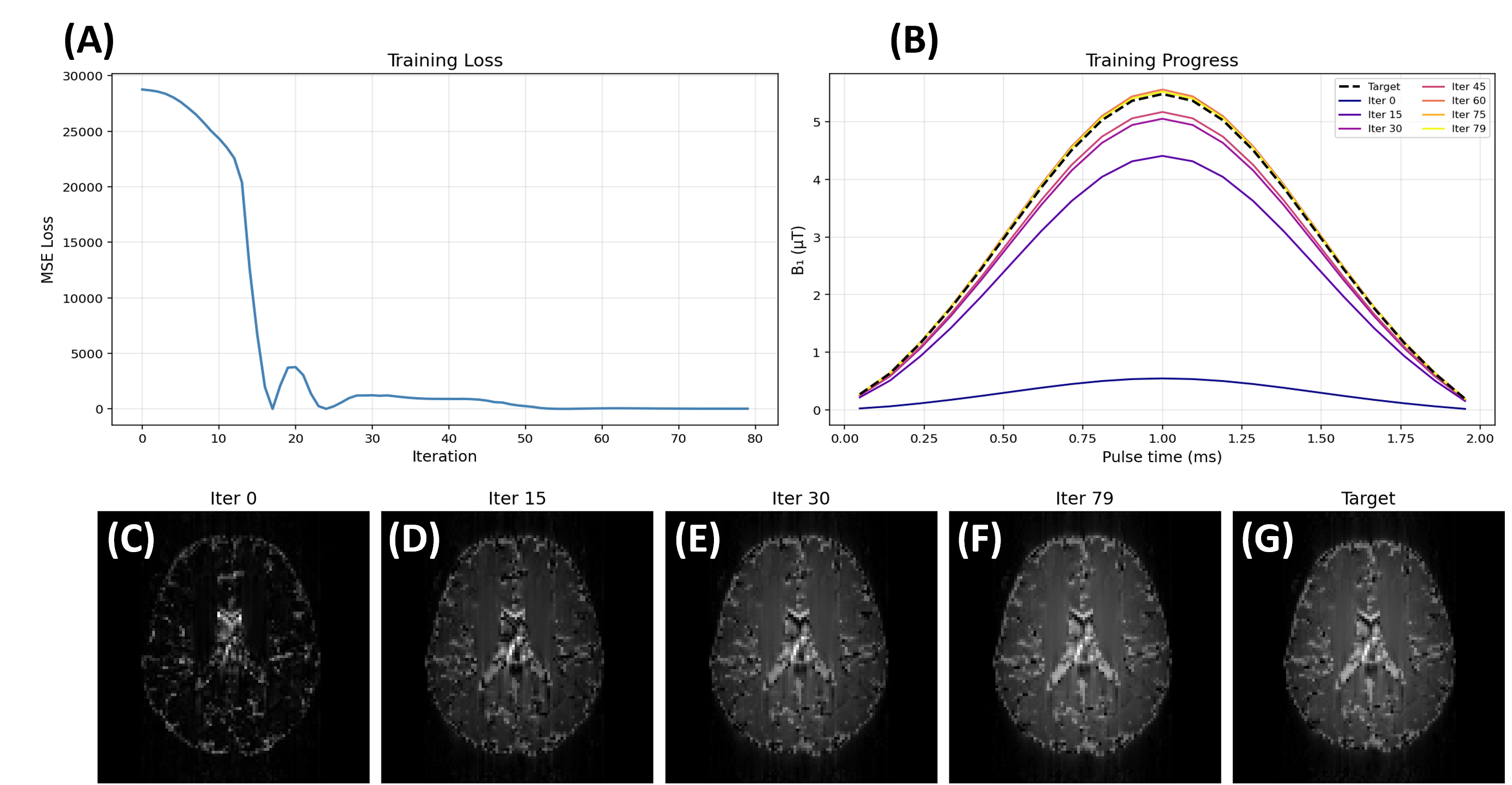}
    \caption{Pulse optimization results for a centric-reordered single-shot TSE sequence ($\text{FA}_\mathrm{ex} = 90\,°$, $\text{FA}_\mathrm{ref} = 150\,°$, $\text{ESP} = 11.1\,\mathrm{ms}$, $\text{ETL} = 100$). (A) MSE loss over 80 iterations. (B) Pulse profile across 21 sub-pulses at selected iterations. (C–F) Reconstructed brain images at iteration 0, 15, 30 and final iteration, and the target reconstruction (G), respectively.}
    \label{fig:opt_sinc}
\end{figure}

To evaluate the differentiability of the approach, we optimized the reference voltage of a sinc-shaped excitation pulse using the image of a 90-degree excited single-shot TSE as a target image. 
\autoref{fig:opt_sinc} shows that gradient-descent-based optimization is able to alter the shaped slice-selective pulse and reach the target with minimal deviation within 80 iterations. 

\section{Discussion}

The off-resonant PDG framework was developed to provide a unified description of RF-driven magnetization dynamics under off-resonant conditions, be it B0 inhomogeneities, off-resonant RF frequency, or gradient-induced off-resonance. The presented examples of spectral selectivity, transverse phase evolution, and spatial slice selection can all be interpreted as different manifestations of a unified frequency-domain RF operator acting on the magnetization system. Native differentiability of the PDG framework provides the possibility for end-to-end optimization by gradient descent methods. Optimization of individual pulses in the context of a sequence's distinct echo-pathways was shown for a TSE pulse optimization using a diferentiable slice-selective EPG approach \cite{augelli2026}. 
The native Pulseq \cite{Pulseq} integration of the framework directly allows for the simulation and optimization of scanner executable sequences, developed within an open source and vendor independent framework.  \\

\noindent
Existing MRI simulation approaches are predominantly based on numerical integration of the Bloch equations using isochromat-based spin models. Frameworks such as KomaMRI \cite{KomaMRI} and JEMRIS \cite{JEMRIS} explicitly simulate the time evolution of a large ensemble of spins under the influence of RF and gradient fields. While this formulation naturally accounts for off-resonant excitation effects and slice selection, through the effective magnetic field, accurate signal formation typically requires a high density of isochromats per voxel, particularly in the presence of spoiler gradients or strong dephasing. This leads to substantial computational cost, especially for high-resolution or large-scale sequence simulations. An alternative class of methods is based on the Extended Phase Graph (EPG) formalism, which reduces computational complexity by representing magnetization states in a discrete coherence domain. Variants such as slice-selective EPG (ssEPG) \cite{Ostenson2020} and spinor-based EPG models \cite{Li2024SpinorEPG} have extended the framework to incorporate slice-selective excitation and more complex RF behavior. However, these approaches generally rely on assumptions of periodicity or structured sequence evolution, which limit their applicability to non-repetitive or strongly transient imaging regimes. Hybrid Bloch–EPG formulations \cite{Guenthner2021UnifyingEPG} attempt to bridge these two paradigms by combining accurate Bloch-based RF modeling with the computational efficiency of EPG state tracking. Despite these advances, such methods still inherit structural constraints from the underlying EPG representation, particularly the requirement for coherence ordering and sequence regularity.

\noindent
In contrast, the extended PDG framework presented in this work does not require periodic sequence structure or voxel-wise isochromat discretization. By formulating RF–magnetization interactions in a generalized operator framework, it enables consistent treatment of off-resonant excitation, slice selection, and spectral effects without relying on coherence-state truncation or large spin ensembles. This provides a more flexible basis for simulating arbitrary pulse sequences. While subdivision of shaped pulses into a series of block pulses is not a new concept and should also work with conventional on-resonant PDG when off-resonant dephasing is modeled between RF pulses, efficiency and accuracy is superior when the sub-pulses incorporate the off-resonant pulse response. Relaxation during RF irradiation with respect to T1, T2, and T2* is modeled accurately as long as enough pulse samples are used. Magnetization transfer effects are still missing in the current formalism; thus, T2-related inversion inefficiency can be modeled, but not yet MT-related inefficiency related to $T2_{MT}$ or the short T1 component of the water--MT system. Likewise, $T1_{\rho}$ and $T2_{\rho}$ effects beyond T1--T2 mixtures are not yet described. Yet, extensions for EPG regarding exchange \cite{Malik2018EPGX} should translate to PDG.
\subsubsection*{Related PDG work}
The PDG extension for off-resonant pulse response was first presented by Dietz et al. at ESMRMB 2025 \cite{Dietz2025ESMRMB}. Duarte et al. demonstrated its application in breast tissue \cite{duarte2025breast}. The modelling of shaped pulses and slice-selective excitation has been accepted for presentation at the 2026 annual meeting of the DACH-ISMRM \cite{Dietz2026DSISMRM}. Recent work by Hussain et al. \cite{hussain2026ssPDG} introduces a different approach to extend phase distribution graphs for slice-selective functionality.

\section{Conclusions}
Extending the Phase Distribution Graph framework for off-resonant pulse response allows simulation of several key MRI features, such as fat--water separation and quantitative mapping of $B_0$ and $B_1$ inhomogeneities using WASABI and Bloch-Siegert sequences. RF discretization into individual instant off-resonant events allows efficient and accurate modeling of the frequency response of arbitrary RF shapes and slice-selective excitation, including related artifacts such as fat shifts or in-flow. Retaining the inherent differentiability of the PDG simulation framework enables optimization of sequences in the frequency domain.

\end{multicols}



%



\begin{multicols}{2}
\bibliography{references} 

@article{Endres2024,
author             = {Endres, Jonathan and Weinm\"uller, Simon and Dang, N. and Zaiss, Moritz},
doi                = {10.1002/mrm.30055},
eprint             = {https://onlinelibrary.wiley.com/doi/pdf/10.1002/mrm.30055},
journal            = {MRM},
number             = {3},
pages              = {1189-1204},
title              = {Phase distribution graphs for fast, differentiable, and spatially encoded Bloch simulations of arbitrary MRI sequences},
url                = {https://onlinelibrary.wiley.com/doi/abs/10.1002/mrm.30055},
volume             = {92},
year               = {2024}
}

@article{Ostenson2020,
author = {Ostenson, Jason and Smith, David S. and Does, Mark D. and Damon, Bruce M.},
title = {Slice-selective extended phase graphs in gradient-crushed, transient-state free precession sequences: An application to MR fingerprinting},
journal = {Magnetic Resonance in Medicine},
volume = {84},
number = {6},
pages = {3409-3422},
doi = {https://doi.org/10.1002/mrm.28381},
url = {https://onlinelibrary.wiley.com/doi/abs/10.1002/mrm.28381},
eprint = {https://onlinelibrary.wiley.com/doi/pdf/10.1002/mrm.28381},
year = {2020}
}

@article{hennig_multiecho_2003,
 author               = {Hennig, Juergen and Weigel, Matthias and Scheffler, Klaus},
 doi                  = {10.1002/mrm.10391},
 issn                 = {1522-2594},
 journal              = {Magnetic Resonance in Medicine},
 number               = {3},
 pages                = {527--535},
 shorttitle           = {Multiecho sequences with variable refocusing flip angles},
 title                = {Multiecho sequences with variable refocusing flip angles: {Optimization} of signal behavior using smooth transitions between pseudo steady states ({TRAPS})},
 url                  = {https://onlinelibrary.wiley.com/doi/abs/10.1002/mrm.10391},
 urldate              = {2023-07-26},
 volume               = {49},
 year                 = {2003},
 }

@article{layton_pulseq_2017,
 author               = {Layton, Kelvin J. and Kroboth, Stefan and Jia, Feng and Littin, Sebastian and Yu, Huijun and Leupold, Jochen and Nielsen, Jon-Fredrik and Stöcker, Tony and Zaitsev, Maxim},
 doi                  = {10.1002/mrm.26235},
 issn                 = {1522-2594},
 journal              = {Magnetic Resonance in Medicine},
 language             = {en},
 number               = {4},
 pages                = {1544--1552},
 shorttitle           = {Pulseq},
 title                = {Pulseq: {A} rapid and hardware-independent pulse sequence prototyping framework},
 url                  = {https://onlinelibrary.wiley.com/doi/abs/10.1002/mrm.26235},
 urldate              = {2024-04-02},
 volume               = {77},
 year                 = {2017},
 }

@inproceedings{Loktyushin2021,
 author               = {Loktyushin, Alexander and others},
 booktitle            = {Medical Image Computing and Computer Assisted Intervention -- MICCAI 2021},
 doi                  = {10.1007/978-3-030-87209-7_41},
 pages                = {709--724},
 title                = {MR-zero: Learning MRI sequence design from scratch},
 volume               = {12901},
 year                 = {2021},
 }

@article{weigel_extended_2015,
 author               = {Weigel, Matthias},
 doi                  = {10.1002/jmri.24619},
 issn                 = {1522-2586},
 journal              = {Journal of Magnetic Resonance Imaging},
 language             = {en},
 number               = {2},
 pages                = {266--295},
 shorttitle           = {Extended phase graphs},
 title                = {Extended phase graphs: {Dephasing}, {RF} pulses, and echoes - pure and simple},
 url                  = {https://onlinelibrary.wiley.com/doi/abs/10.1002/jmri.24619},
 urldate              = {2023-11-30},
 volume               = {41},
 year                 = {2015},
 }

@unpublished{Dietz2026DSISMRM,
  author       = {Felix Dietz and Simon Weinm{\"u}ller and Jonathan Endres and Moritz Zaiss},
  title        = {Advancing Phase Distribution Graphs: Time-effective simulations including shaped pulses and slice-selective excitation},
  note         = {Accepted for presentation at the DACH-ISMRM Annual Meeting, Bamberg 2026},
  year         = {April 20, 2026},
  month        = {April},
  day          = {20},
  address      = {Bamberg, Germany}
}

@article{WASABI,
author = {Schuenke, Patrick and Windschuh, Johannes and Roeloffs, Volkert and Ladd, Mark E. and Bachert, Peter and Zaiss, Moritz},
title = {Simultaneous mapping of water shift and B1(WASABI)—Application to field-Inhomogeneity correction of CEST MRI data},
journal = {Magnetic Resonance in Medicine},
volume = {77},
number = {2},
pages = {571-580},
doi = {https://doi.org/10.1002/mrm.26133},
url = {https://onlinelibrary.wiley.com/doi/abs/10.1002/mrm.26133},
eprint = {https://onlinelibrary.wiley.com/doi/pdf/10.1002/mrm.26133},
year = {2017}
}

@article{BlochSiegert,
author = {Sacolick, Laura I. and Wiesinger, Florian and Hancu, Ileana and Vogel, Mika W.},
title = {B1 mapping by Bloch-Siegert shift},
journal = {Magnetic Resonance in Medicine},
volume = {63},
number = {5},
pages = {1315-1322},
doi = {https://doi.org/10.1002/mrm.22357},
url = {https://onlinelibrary.wiley.com/doi/abs/10.1002/mrm.22357},
eprint = {https://onlinelibrary.wiley.com/doi/pdf/10.1002/mrm.22357},
year = {2010}
}

@article{Pulseq,
author = {Kelvin J. Layton et. al.},
title = {{Pulseq: A rapid and hardware-independent pulse sequence prototyping framework}},
journal = {Magnetic Resonance in Medicine},
volume = {77},
number = {4},
pages = {1544-1552},
doi = {https://doi.org/10.1002/mrm.26235},
url = {https://onlinelibrary.wiley.com/doi/abs/10.1002/mrm.26235},
eprint = {https://onlinelibrary.wiley.com/doi/pdf/10.1002/mrm.26235},
year = {2017}
}

@inproceedings{Li2024SpinorEPG,
  author    = {Z. Li and J. Zou and C. Liu and R. Li},
  title     = {Spinor-EPG: An Improved EPG Algorithm for Fast Simulation of the Non-Ideal Slice Profile Effect in MRF},
  booktitle = {Proceedings of the 32nd International Society for Magnetic Resonance in Medicine},
  year       = {2024},
  volume      = {2024},
  pages       = {3557}
}

@article{Guenthner2021UnifyingEPG,
  author       = {Guenthner, Christian and Amthor, Thomas and Doneva, Mariya and Kozerke, Sebastian},
  title        = {A Unifying View on Extended Phase Graphs and Bloch Simulations for Quantitative MRI},
  journal      = {Scientific Reports},
  year         = {2021},
  volume       = {11},
  number       = {1},
  pages        = {21289},
  month        = oct,
  doi          = {10.1038/s41598-021-00233-6},
  pmid         = {34711847},
  pmcid        = {PMC8553818},
  issn         = {2045-2322},
  publisher    = {Nature Publishing Group}
}

@article{Haase_CHESS,
  author    = {A. Haase and J. Frahm and W. H{\"a}nicke and D. Matthaei},
  title     = {{1H NMR chemical shift selective (CHESS) imaging}},
  journal   = {Physics in Medicine and Biology},
  volume    = {30},
  number    = {4},
  pages     = {341--344},
  year      = {1985},
  doi       = {10.1088/0031-9155/30/4/008}
}

@article{collins1998brainweb,
  author={Collins, D.L. and Zijdenbos, A.P. and Kollokian, V. and Sled, J.G. and Kabani, N.J. and Holmes, C.J. and Evans, A.C.},
  journal={IEEE Transactions on Medical Imaging}, 
  title={Design and construction of a realistic digital brain phantom}, 
  year={1998},
  volume={17},
  number={3},
  pages={463-468},
  doi={10.1109/42.712135},
  note      = {Available at \url{https://brainweb.bic.mni.mcgill.ca/}}
}

@article{KomaMRI,
  author = {Castillo-Passi, Carlos and Coronado, Ronal and Varela-Mattatall, Gabriel and Alberola-López, Carlos and Botnar, René and Irarrazaval, Pablo},
  title = {KomaMRI.jl: An open-source framework for general MRI simulations with GPU acceleration},
  journal = {Magnetic Resonance in Medicine},
  doi = {https://doi.org/10.1002/mrm.29635},
  url = {https://onlinelibrary.wiley.com/doi/abs/10.1002/mrm.29635},
  eprint = {https://onlinelibrary.wiley.com/doi/pdf/10.1002/mrm.29635},
year = {2023}
}

@article{JEMRIS,
author = {Stöcker, Tony and Vahedipour, Kaveh and Pflugfelder, Daniel and Shah, N. Jon},
title = {High-performance computing MRI simulations},
journal = {Magnetic Resonance in Medicine},
volume = {64},
number = {1},
pages = {186-193},
doi = {https://doi.org/10.1002/mrm.22406},
url = {https://onlinelibrary.wiley.com/doi/abs/10.1002/mrm.22406},
eprint = {https://onlinelibrary.wiley.com/doi/pdf/10.1002/mrm.22406},
year = {2010}
}

@misc{hussain2026ssPDG,
  title        = {Slice-Profile-Enabled Phase Distribution Graphs for MRI Simulation},
  author       = {Snawar Hussain and Daniel C. Hoinkiss and J{\"o}rn Huber and Vincent Kuhlen and Matthias G{\"u}nther},
  year         = {2026},
  howpublished = {\url{https://arxiv.org/abs/2606.09233}},
  note         = {arXiv:2606.09233 [physics.med-ph]}
}

@article{augelli2026,
author = {Augelli, Madison M. and Sharma, Anuj and Griswold, Mark A. and Grissom, William A.},
title = {Optimizing Selective RF Pulses for Enhanced Signal Stability in Turbo Spin Echo Using a Differentiable Extended Phase Graph Model},
journal = {Magnetic Resonance in Medicine},
volume = {96},
number = {1},
pages = {214-226},
doi = {https://doi.org/10.1002/mrm.70340},
url = {https://onlinelibrary.wiley.com/doi/abs/10.1002/mrm.70340},
eprint = {https://onlinelibrary.wiley.com/doi/pdf/10.1002/mrm.70340},
year = {2026}
}

@misc{mrzero_framework,
  author       = {{MR-zero Development Team}},
  title        = {MR-zero MR simulation framework},
  howpublished = {\url{https://mrsources.github.io/MRzero-Core/}},
  year         = {2021},
  note         = {Accessed: 2026-06-13}
}

@inproceedings{duarte2025breast,
  author    = {Magda Duarte and Felix Dietz and Tobias Dornstetter and Jonathan Endres and Simon Weinm{\"u}ller and Sebastian Bickelhaupt and Moritz Zaiss},
  title     = {Breast Digital Twin: simulating with {MR-zero}},
  booktitle = {Book of Abstracts ESMRMB 2025 Online 41st Annual Scientific Meeting 8--11 October 2025},
  series    = {Magnetic Resonance Materials in Physics, Biology and Medicine},
  volume    = {38},
  number    = {Suppl. 1},
  pages     = {055},
  year      = {2025},
  doi       = {10.1007/s10334-025-01278-8},
  note      = {ESMRMB 2025, Marseille, France, October 8--11, 2025}
}

@inproceedings{Dietz2025ESMRMB,
  author    = {Felix Dietz and Simon Weinm{\"u}ller and Jonathan Endres and Moritz Zaiss},
  title     = {Phase graph-based {MRI} simulation including off-resonant pulse response},
  booktitle = {Book of Abstracts ESMRMB 2025 Online 41st Annual Scientific Meeting 8--11 October 2025},
  series    = {Magnetic Resonance Materials in Physics, Biology and Medicine},
  volume    = {38},
  number    = {Suppl. 1},
  pages     = {130},
  year      = {2025},
  doi       = {10.1007/s10334-025-01278-8},
  note      = {ESMRMB 2025, Marseille, France, October 8--11, 2025}
}

@article{Malik2018EPGX,
author = {Malik, Shaihan J. and Teixeira, Rui Pedro A.G. and Hajnal, Joseph V.},
title = {Extended phase graph formalism for systems with magnetization transfer and exchange},
journal = {Magnetic Resonance in Medicine},
volume = {80},
number = {2},
pages = {767-779},
doi = {https://doi.org/10.1002/mrm.27040},
url = {https://onlinelibrary.wiley.com/doi/abs/10.1002/mrm.27040},
eprint = {https://onlinelibrary.wiley.com/doi/pdf/10.1002/mrm.27040},
year = {2018}
}
\end{multicols}

\section*{Figures}

\end{document}